\documentclass{bmcart}

\usepackage{amsthm,amsmath}
\usepackage{braket}
\usepackage[utf8]{inputenc} 
\usepackage{graphicx}

\startlocaldefs
\endlocaldefs

\begin{document}

\begin{frontmatter}

\begin{fmbox}
\dochead{Research}


\title{Digital quantum simulation of quantum gravitational entanglement with IBM quantum computers}


\author[
  addressref={aff1},                   
  corref={aff1},                       
  email={carlos.sabin@uam.es}   
]{\inits{C.}\fnm{Carlos} \snm{Sabín}}
%

\address[id=aff1]{
  \orgdiv{Departamento de Física Teórica},             
  \orgname{Universidad Autónoma de Madrid},          
 \postcode{28049},
  \city{Madrid},                              
  \cny{Spain}                                    
}




\end{fmbox}


\begin{abstractbox}

\begin{abstract} 

We report the digital quantum simulation of a hamiltonian involved in the generation of quantum entanglement by gravitational means. In particular, we focus on a pair of quantum harmonic oscillators, whose interaction via a quantum gravitational field generates single-mode squeezing in both modes at the same time, a non-standard process in quantum optics. We perform a boson-qubit mapping and a digital gate decomposition specific for IBM quantum devices. We use error mitigation and post-selection to achieve high-fidelity, accessing a parameter regime out of direct experimental reach.

\end{abstract}




\end{abstractbox}
%

\end{frontmatter}



\section*{Introduction}

Experimental tests of a quantum theory of gravity are entirely beyond experimental reach. A more modest approach has emerged recently, where the focus is just to prove the quantum nature of gravity, even without unveiling the underlying full quantum theory \cite{Bose:2017nin,Marshman:2019sne,Marletto,interf,newton,versus,constraints,julen}. The idea behind this novel approach is that if we are able to generate quantum entanglement between quantum systems by gravitational means, then we know that gravity must be quantum. While these experimental proposals are still beyond the reach of state-of-the-art quantum technology, it is at least conceivable that the rapid advancement of the experimental setups could change this scenario in the short/medium term, enabling a direct test of the quantum side of gravity.

In \cite{PhysRevD.105.106028}, the mechanism of entanglement generation by a quantum gravitational field is elucidated, by considering a hamiltonian formalism and a setup of two quantum harmonic oscillators in the presence of a gravitational field. The quantum treatment of the field gives rise to interaction hamiltonians between the two bosonic modes, which evolve from the ground state to an entangled one. The authors distinguish between two different regimes, static and non-static. In the latter, the hamiltonian generates single-mode squeezing in the two modes at the same time, a non-standard interaction in conventional quantum optics, by means of the exchange of a graviton, which is then the entanglement source. Then, the detection of the evolution from a ground state to an entangled state with a small probability amplitude of containing excitation pairs in both modes would be the signature of the quantum nature of the field. However, the parameter regime at which this interaction would be dominant lies beyond current experimental reach \cite{PhysRevD.105.106028}. In this paper, we address this issue, by considering a quantum platform in which these results can be simulated. The non-standard nature of the interaction makes it difficult to think of an analogue quantum simulation, so we will consider a digital approach.

We have recently introduced a recipe for digital quantum simulation of multimode bosonic hamiltonians \cite{sabindqs}. It consists in the combination of Trotter \cite{lloyd} and gate-decomposition techniques \cite{sommalallama} with a boson-qubit mapping \cite{losalamos,sommathesis} in order to encode bosonic hamiltonians into a sequence of single-qubit and two-qubit gates. We have applied this scheme to the high-fidelity digital quantum simulation of paradigmatic quantum-optics interactions, such as beam-splitters or single-mode and two-mode squeezers \cite{paula}.

In this work, we apply this scheme to the digital quantum simulation in IBM quantum devices of the quantum gravitational hamiltonian of \cite{PhysRevD.105.106028}. We translate the bosonic two-mode hamiltonian into a multiqubit interaction, which is then decomposed into the gate set available in IBM quantum computers. In order to show the potential of current quantum technology, we consider the simplest scenario, where the maximum number of excitations allowed per mode is restricted to 2 and the fidelity is approximated by the leading order in a perturbative approach. Moreover, we make use of error mitigation and post-selection techniques to achieve high-fidelity quantum simulations, with  fidelities above 90 \%. These preliminary few-qubit experiments can be understood as a first step towards larger quantum simulations including more modes and more photons per mode, which would go beyond the capabilities of classical computers. In this post-classical regime, digital quantum simulations would be a unique way of informing the developments of the actual quantum gravitational experiments. Moreover, the generation of an entangled state in such a setup could be considered an indirect proof --or, at least, a reassuring argument-- of the generation of quantum entanglement by gravitational means, and therefore, of the quantum nature of gravity.

The structure of the paper is the following. In the next section, we introduce the hamiltonian of interest and apply the digitization framework to it. Then, we report on the results of the fidelity for the circuits of section II after launching them in a suitable quantum computer. In the last section we summarize our results and discuss extensions and generalizations.

\section*{Digitization}
 
Following \cite{PhysRevD.105.106028}, we consider two quantum harmonic oscillators interacting via a gravitational field. While a consistent full quantum treatment of the field remains of course unknown, we can always adopt an effective field approach, in which a small linear perturbation to the Minkowski metric is promoted to a quantum field in terms of spin-2 graviton creation and annihilation operators. In so doing, in \cite{PhysRevD.105.106028} it is shown that if the harmonic oscillators are in fixed positions -static regime- then the leading interaction terms between them are two-mode squeezing and beam-splitter hamiltonians. This class of hamiltionians was already digitized in \cite{sabindqs, paula}, giving rise to high-fidelity quantum simulations.
 
 If the oscillators are non-static, then a new 
 Hamiltonian \cite{PhysRevD.105.106028} appears
 \begin{equation}
     H_{AB}= -\hbar g \left(a^{\dagger 2}b^{\dagger 2}+ a^{2}b^{2}\right)\label{eq:ham}
 \end{equation}
where A and B are two harmonic oscillators separated by a distance $d/2$, with frequency $\omega_m$ and creation operators $a^{\dagger}$, $b^{\dagger}$, respectively, and $\hbar$ is the reduced Planck constant, as usual. The coupling between the oscillators is 
\begin{equation}
    g=\frac{9 G\hbar \omega_m^2}{16 c^4 d},\label{eq:coupling}
\end{equation}
$G$ being Newton constant and $c$ the speed of light in vacuum, as usual. 
This hamiltonian would dominate the entanglement generation in a regime of frequencies that is beyond experimental reach $\omega_m>10^{21} Hz$ \cite{PhysRevD.105.106028}. Interestingly, this double single-mode squeezing interaction is not usually discussed in standard quantum optics scenarios, as highlighted in \cite{PhysRevD.105.106028}. 

Notice that the hamiltonian in Eq.(\ref{eq:ham}) can only exist if the field is quantum and that regardless the actual underlying quantum theory, it would account for the quantum gravitational field as long as it can be treated as a linear perturbation, which is of course the case for weak gravitational fields, such as  the one generated by the Earth near its surface. Therefore, detecting the effects of the physical action of Eq. (\ref{eq:ham}) would amount to detecting the quantum nature of the gravitational field or, in other words, the existence of gravitons. In particular, if the initial state is the ground state of the two quantum harmonic oscillators, $|0\rangle_A\,|0\rangle_B$, the evolution under Eq. (\ref{eq:ham}) would give rise to a state \cite{PhysRevD.105.106028}:
\begin{equation}
   |\Psi\rangle=\frac{1}{\sqrt{1+(g/(2\omega_m)^2)}} \left(|0\rangle_A\,|0\rangle_B+\frac{g}{2\,\omega_m}|2\rangle_A\,|2\rangle_B\right),\label{eq:coupling}
\end{equation}
containing a small probability of having two excitations per mode. This is an entangled state, with concurrence \cite{PhysRevD.105.106028}:
\begin{equation}
C\simeq\frac{\sqrt 2\, g}{\omega_m}.
\end{equation}

Thus, proving the evolution from the ground state to a state with a certain probability amplitude of $|2\rangle_A\,|2\rangle_B$ would prove the generation of quantum entanglement by gravitational means and therefore the quantum nature of the gravitational field.

Therefore, we want to simulate the unitary
\begin{equation}
U_\epsilon=e^{-iH_{AB} t/\hbar}=e^{i\varepsilon \left(a^{\dagger2}b^{\dagger2}+a^{2}b^{2}\right)},\label{eq:unitary}
\end{equation}
where 
\begin{equation}
\varepsilon=g\, t. \label{eq:epsilon}
\end{equation}
For $\omega_m=10^{21} Hz$ and $d=10^{-4} m$, we would have $g\approx 10^{-31} Hz$. However, in our simulated scenario we could consider even larger values of $\omega_m$, leveraging the fact that the coupling grows quadratically with $\omega_m$. This would give rise also to larger entanglement generation.

The main goal of this paper is the digitization of the unitary in Eq. (\ref{eq:unitary}). For this, we start with a boson-qubit mapping. As shown in \cite{losalamos, sommathesis}, it is possible to map N bosonic modes containing a maximum number of $N_p$ excitations each to $N(N_p+1)$ qubits. In our current case, we have to consider only two modes, and we can start by restricting ourselves to two maximum excitations, since we can consider an initial ground state and the leading order contribution of Eq. (\ref{eq:unitary}) acting on it. Therefore we will need six qubits, labeled from 0 to 5.  According to the boson-qubit mapping in \cite{losalamos,sommathesis}, we have the following Fock states  $\ket{n}$:
\begin{eqnarray}
\label{bosonmap}
|0\rangle_A &\leftrightarrow& |0_0 1_1 1_2 \rangle \nonumber \\
|1\rangle_A &\leftrightarrow& |1_0 0_1 1_2 \rangle \nonumber \\
|2\rangle_A &\leftrightarrow& |1_0 1_1 0_2  \rangle,
\end{eqnarray} 
for mode A and similarly
\begin{eqnarray}
\label{bosonmap}
|0\rangle_B &\leftrightarrow& |0_3 1_4 1_5 \rangle \nonumber \\
|1\rangle_B &\leftrightarrow& |1_3 0_4 1_5 \rangle \nonumber \\
|2\rangle_B &\leftrightarrow& |1_3 1_4 0_5  \rangle,
\end{eqnarray} 
for mode B, where $|0_i \rangle$, $|1_i \rangle$ ($i=0, 1... 5$) are the states of qubit $i$, which are the eigenstates associated to the positive and negative eigenvalues of the Pauli operator $\sigma_z^i$, respectively.
Using the boson-qubit operator mapping for $N_P=2$,  we can write the bosonic creation operators as:
\begin{eqnarray}
\label{bosonmap2}
a^{\dagger} &\rightarrow &
 \sigma_-^{0}
\sigma_+^{1}+\sqrt{2} \ \sigma_-^{1}
\sigma_+^{2}\nonumber\\b^{\dagger} &\rightarrow &
 \sigma_-^{3}
\sigma_+^{4}+\sqrt{2} \ \sigma_-^{4}
\sigma_+^{5}
\end{eqnarray}
where the Pauli creation and annihilation
operators are given by 
\begin{equation}\label{eq:qubitpm}
\sigma_\pm^k=\frac{1}{2} (\sigma_x^k\pm i\sigma_y^k),
\end{equation}
in terms of the Pauli matrices $\sigma_x$ and $\sigma_y$ ($k=0,1...5$). Notice that, for each qubit $\sigma_+ |0\rangle=0$, $\sigma_- |0 \rangle=|1 \rangle$, $\sigma_- |1\rangle=0$, $\sigma_+ |1 \rangle=|0 \rangle$. 

Using Eq. (\ref{bosonmap2}) and the properties of the Pauli operators:
\begin{eqnarray}
\sigma_{\pm}^k\sigma_{\mp}^k&=&\frac{1}{2}(1\pm\sigma_z^k)\nonumber\\
\sigma_{\pm}^k\sigma_{\pm}^k&=&0,
\end{eqnarray}
we get
\begin{eqnarray}
a^{\dagger2}&\rightarrow&\sqrt{2} \sigma_-^0\sigma_+^2\nonumber\\
a^2&\rightarrow&\sqrt{2} \sigma_+^0\sigma_-^2\nonumber\\
b^{\dagger2}&\rightarrow&\sqrt{2} \sigma_-^3\sigma_+^5\nonumber\\
b^2&\rightarrow&\sqrt{2} \sigma_+^3\sigma_-^5.
\end{eqnarray}
Therefore, we have:
\begin{equation}
a^{\dagger2}b^{\dagger2}+a^{2}b^{2}\rightarrow 2(\sigma_-^0\sigma_+^2\sigma_-^3\sigma_+^5+\sigma_+^0\sigma_-^2\sigma_+^3\sigma_-^5)
\end{equation}
Then, using Eq. (\ref{eq:qubitpm}) and summing up, we get:  
\begin{eqnarray}\label{eq:mappedbs}
a^{\dagger2}b^{\dagger2}+a^{2}b^{2} &\rightarrow&\frac{1}{4}(\sigma_x^{0}\sigma_x^{2}\sigma_x^{3}\sigma_x^{5}+\sigma_x^{0}\sigma_x^{2}\sigma_y^{3}\sigma_y^{5}-\sigma_x^{0}\sigma_y^{2}\sigma_x^{3}\sigma_y^{5}+\sigma_x^{0}\sigma_y^{2}\sigma_y^{3}\sigma_x^{5}+\sigma_y^{0}\sigma_x^{2}\sigma_x^{3}\sigma_y^{5}-\nonumber\\ &&\sigma_y^{0}\sigma_x^{2}\sigma_y^{3}\sigma_x^{5}+\sigma_y^{0}\sigma_y^{2}\sigma_x^{3}\sigma_x^{5}+\sigma_y^{0}\sigma_y^{2}\sigma_y^{3}\sigma_y^{5}).
\end{eqnarray}
Although it might look that the different terms in Eq. (\ref{eq:mappedbs}) do not commute, we can check that they do. Let us consider two terms, and notice that, using the properties of the commutators and the fact that the operators acting on different qubits obviously commute, we have:
\begin{eqnarray}\label{eq:commutators}
[\sigma_x^{0}\sigma_x^{2}\sigma_x^{3}\sigma_x^{5}, \sigma_x^{0}\sigma_y^{2}\sigma_y^{3}\sigma_x^{5}]=[\sigma_x^{0}\sigma_x^{2},\sigma_x^{0}\sigma_y^{2}]\sigma_x^{3}\sigma_x^{5}\sigma_y^{3}\sigma_x^{5}+ \sigma_x^{0}\sigma_y^{2}\sigma_x^{0}\sigma_x^{2}[\sigma_x^{3}\sigma_x^{5},\sigma_y^{3}\sigma_x^{5}]
\end{eqnarray}
Then, similarly:
\begin{eqnarray}
[\sigma_x^{0}\sigma_x^{2},\sigma_x^{0}\sigma_y^{2}]=(\sigma_x^{0})^2
[\sigma_x^{2},\sigma_y^{2}] \end{eqnarray}
\begin{eqnarray}
 [\sigma_x^{3}\sigma_x^{5},\sigma_y^{3}\sigma_x^{5}] =[\sigma_x^{3},\sigma_y^{3}](\sigma_x^{5})^2 \end{eqnarray}
Putting everything together and using the properties of Pauli matrices
$\sigma_i^{(k)}\sigma_j^{(k)}=i\epsilon_{ijk}\sigma_k^{(k)}$ ($\epsilon_{ijk}$ being the Levi-Civita tensor), the two contributions in Eq. (\ref{eq:commutators}) cancel out. And we can proceed similarly with all the terms in Eq. (\ref{eq:mappedbs}).

Since everything commutes, we have 
\begin{equation}
U_{\varepsilon}=\prod_{i=1}^8 U_{\varepsilon}^{(i)},
\end{equation}
where the $ U_{\varepsilon}^{(i)}$'s are given by Eq. (\ref{eq:mappedbs}). For instance:
\begin{equation}\label{eq:mappedbsu1}
U_{\varepsilon}^{(1)}=e^{i\frac{\varepsilon}{4}\sigma_x^{(0)}\sigma_x^{(2)}\sigma_x^{(3)}\sigma_x^{(5)}}.
\end{equation}
However, these unitaries are not directly available in the quantum devices of IBM. Therefore, we have to perform a gate decomposition to express it in terms of the desired gate set. 
In general, if we find an unitary operation $U$ such that:
\begin{equation}\label{eq:gatedec1}
H=U^{\dagger}H_0 U,
\end{equation}
then we can write the dynamics governed by the Hamiltonian $H$ as:
\begin{equation}\label{eq:gatedec2}
e^{i H \varepsilon}=U^{\dagger}e^{i H_0 \varepsilon} U.
\end{equation}

\begin{figure*}[h!]
   \includegraphics[width=0.8\textwidth]{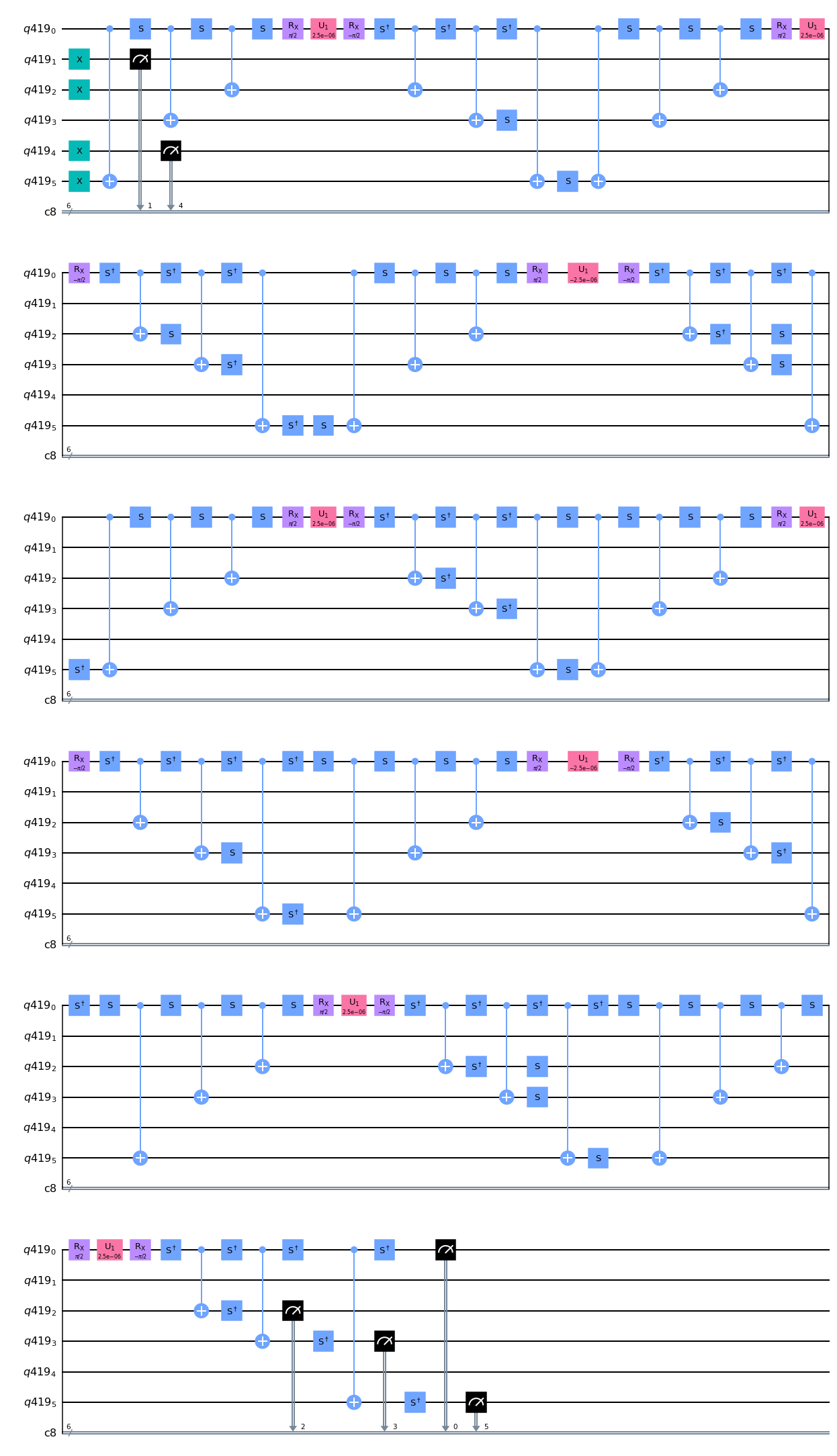}
    \caption{Quantum circuit for the unitary Eq. (\ref{eq:unitarysim}) and $\varepsilon=0.05\cdot10^{-6}$.}
    \label{fig:1} 
\end{figure*}

\begin{figure*}[h!]
    \includegraphics[width=0.8\textwidth]{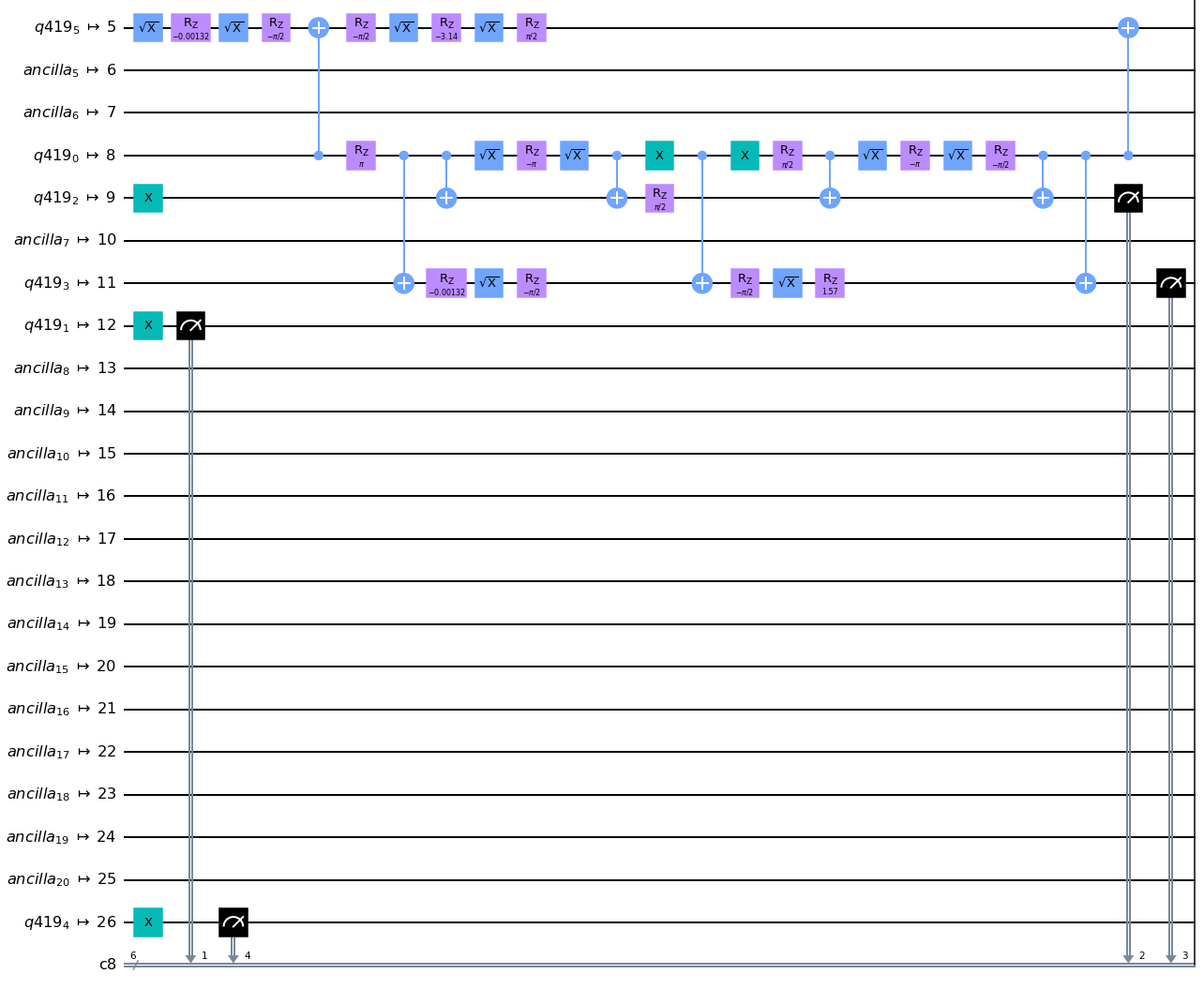}
    \caption{Final transpiled circuit run in Auckland for the same circuit of \ref{fig:1}.}
    \label{fig:2} 
\end{figure*}

Thus we can perform a separate gate decomposition for each $U_{\varepsilon}^{(i)}$. For instance, for the first term we can use:
\begin{eqnarray}
e^{-i\frac{\pi}{4}\sigma_x^{0}}(-\sigma_z^{0})e^{i\frac{\pi}{4}\sigma_x^{0}}&=&\sigma_y^{0}\\\nonumber
e^{-i\frac{\pi}{4}\sigma_z^{0}\sigma_x^{2}}\sigma_y^{0}e^{i\frac{\pi}{4}\sigma_z^{0}\sigma_x^{2}}&=&-\sigma_x^{0}\sigma_x^{2}\\\nonumber
e^{-i\frac{\pi}{4}\sigma_z^{0}\sigma_x^{3}}(-\sigma_x^{0}\sigma_x^{2})e^{i\frac{\pi}{4}\sigma_z^{0}\sigma_x^{3}}&=&-\sigma_y^{0}\sigma_x^{2}\sigma_x^{3}\\\nonumber
e^{-i\frac{\pi}{4}\sigma_z^{0}\sigma_x^{5}}(-\sigma_y^{0}\sigma_x^{2}\sigma_x^{3})e^{i\frac{\pi}{4}\sigma_z^{0}\sigma_x^{5}}&=&\sigma_x^{0}\sigma_x^{2}\sigma_x^{3}\sigma_x^{5}.
\end{eqnarray}
Therefore, by using Eqs. (\ref{eq:gatedec1}),(\ref{eq:gatedec2}) and (\ref{eq:mappedbsu1}) we find:
\begin{equation}
U_{\varepsilon}^{(1)}=U^\dagger e^{-i\frac{\varepsilon}{4}\sigma_z^{0}}U,\label{eq:upm1}
\end{equation}
where:
\begin{equation}
U= e^{i\frac{\pi}{4}\sigma_x^{0}}e^{i\frac{\pi}{4}\sigma_z^{0}\sigma_x^{2}}e^{i\frac{\pi}{4}\sigma_z^{0}\sigma_x^{3}}e^{i\frac{\pi}{4}\sigma_z^{0}\sigma_x^{5}}.\label{eq:u1}
\end{equation}
Note that similar decompositions can be obtained for the rest of $ U_{\epsilon}^{(i)}$'s by adding at the end of the string the number of $e^{i\pi/4\sigma_z^{(i)}}$ necessary to rotate some of the $\sigma_x$ to $\sigma_y$ --although there might be more efficient decompositions for each particular case. With this procedure we will have a total number of 24 single-qubit rotations and 24 two-qubit $ZX$-gates. An extra step is needed though, which is the conversion of the ZX gates into the CNOT gates available in the IBM architecture. The two-qubit gates can be translated into CNOT gates by using:
\begin{equation}
e^{i\frac{\pi}{4}\sigma_z^{0}\sigma_x^{j}}= e^{i\frac{\pi}{4}\sigma_x^j}e^{i\frac{\pi}{4}\sigma_z^{0}}e^{-i\frac{\pi}{4}} CNOT^{0-j} \label{eq:cnot},
\end{equation}
where the CNOT gate between a pair of qubits $i$, $j$ is defined as:
\begin{equation}
CNOT^{i-j}= \begin{pmatrix}1&0 &0 &0\\0&0&0 &1 \\ 0&0 &1 &0 \\ 0&1 &0 &0\end{pmatrix}.
\end{equation}

We also define the single-qubit single-parameter rotation
\begin{equation}
U_1(\lambda)= \begin{pmatrix}1&0\\0&e^{i\lambda}\end{pmatrix}=e^{\frac{i\lambda}{2}}e^{-\frac{i\lambda\sigma_z}{2}},\label{eq:sqg1}
\end{equation}
the single-qubit S gate,
\begin{equation}
S= \begin{pmatrix}1&0\\0&i\end{pmatrix}=e^{i\frac{\pi}{4}}e^{-i\frac{\pi}{4}\sigma_z},\label{eq:sqg2}
\end{equation}
and the single- qubit $R_X(\theta)$ gate:
\begin{equation}
R_X(\theta)= e^{-i\frac{\theta X}{2}},\label{eq:sqg2}
\end{equation}
where $X=\sigma_x$, using the same conventions as IBM.

Putting everything together, some operations on qubit 0 cancel out and we get -up to global phases- a final expression with 9 single-qubit gates and six CNOT gates:

\begin{eqnarray} 
U_{\varepsilon}^{(1)}&=& CNOT^{0-5}S^0CNOT^{0-3}S^0CNOT^{0-2}S^0R_X^0\left(\frac{\pi}{2}\right)U_1\left(\frac{\varepsilon}{2}\right)R_X^0\left(\frac{\pi}{2}\right)S^{\dagger 0}\nonumber\\& & CNOT^{0-2}S^{\dagger 0}CNOT^{0-3}S^{\dagger 0}CNOT^{0-5}.\label{eq:unitarysim}
\end{eqnarray}

As mentioned above, the rest of the $ U_{\varepsilon}^{(i)}$'s would be obtained by adding the necessary $S$ gates at the end of the string.

In Figure 1, we see the circuit corresponding to the whole unitary $U_{\varepsilon}$ for $\varepsilon=0.5\cdot10^{-6}$ acting on an initial ground state as launched by us in IBM Auckland and in Figure 2 the final ``transpiled'' version.  All the circuits were transpiled using qiskit optimization level 3 \cite{qiskit}. Note that the initial four X-gates in Figure 1  are required for the preparation of the initial ground state 
\begin{equation}
    |0\rangle_A |0\rangle_B\leftrightarrow |0_0 1_1 1_2 \rangle |0_3 1_4 1_5 \rangle 
\end{equation}
and that in Figure 2, the definitions are:
\begin{equation}
\sqrt{X}= \frac{1}{2}\begin{pmatrix}1+i&1-i\\1-i&1+1\end{pmatrix}=e^{i\frac{\pi}{4}}R_X\left(\frac{\pi}{2}\right),
\end{equation}
and
\begin{equation}
    R_Z(\theta)=e^{-\frac{i\theta Z}{2}},
\end{equation}
with $Z=\sigma_z$.

As can be seen in Figure 2, the transpilation simplifies the circuit significantly, since the whole unitary only comprises 9 CNOT gates and 30 single-qubit gates.

\section*{Fidelity}
In order to assess the performance of our digitization, the goal now is to characterize the fidelity of the state. Notice that the restriction to a maximum of two photons per mode corresponds to a restriction to perturbative values of $\varepsilon$. In second-order perturbation theory, the state that we intend to simulate would have the form 
\begin{equation}\label{eq:statepert}
|\psi\rangle=(1-\varepsilon^2)|0\rangle_A|0\rangle_B-i\,\sqrt{2}\varepsilon|2\rangle_A|2\rangle_B.
\end{equation}

Our aim is to compute the fidelity \cite{nielsenchuang} 
\begin{eqnarray}
    F(\left\lvert \psi\right\rangle,\rho)=\left\langle\psi\right\rvert\rho\left\lvert\psi\right\rangle.
\end{eqnarray}
where $\psi$ is the aforementioned perturbative state and $\rho$ is the state actually obtained in the experiment. 
 The leading order term of the fidelity would be 
\begin{equation}\label{eq:fidapprox}
F=P_0, 
\end{equation}
where $P_0$ is the probability of the ground state in the state $\rho$, which can be easily retrieved from the experiment with qiskit, thus obtaining a second-order approximation error. Moreover,  we are only interested in the initial ground state and in the dynamics generated by the hamiltonian (\ref{eq:ham}), so we can also use post-selection, by simply neglecting all the probability counts of the states that are not $|0\rangle_A |0\rangle_B\rightarrow|0_0 1_1 1_2 \rangle |0_3 1_4 1_5 \rangle$ or $|2\rangle_A |2\rangle_B\rightarrow|1_0 1_1 0_2 \rangle |1_3 1_4 0_5 \rangle$.
We also make use of error mitigation techniques in the read-out process \cite{mitigation}. 

We will present results in IBM Auckland,  a 27-qubit machine which displays both a high quantum volume $\operatorname{QV}=64$ \cite{quantumvolume} and the connectivity required for our purpose, that is, the availability of the CNOT gates needed to implement the unitary. 
Typical average error rates in Auckland are $1.127\times 10^{-2}$ for readout (note however that we have used error mitigation for the readout), $4.278\times 10^{-4}$ for single-qubit gates and $1.413\times 10^{-2}$ for CNOT gates.  Error bars can be assigned by considering a typical average readout assignment error and standard error propagation techniques. Note however that all these parameters change on a daily basis. We realized the experiments on March 2, 2022.
With 9 CNOT gates, 30 single-qubit gates and six readouts and the error rates above fidelity would be expected to drop below 80 $\%$. However, with the mentioned use of post-selection and mitigation we have been able to retrieve fidelities close to 1, as can be seen in Figure 3. The fraction of discarded measurements in the post-selection measurements is around 10 \%.
Note that we are considering a parameter range that, while still perturbatively small, goes beyond what can be achieved experimentally, as discussed above. 
\begin{figure*}[h!]
\includegraphics[width=0.8\textwidth]{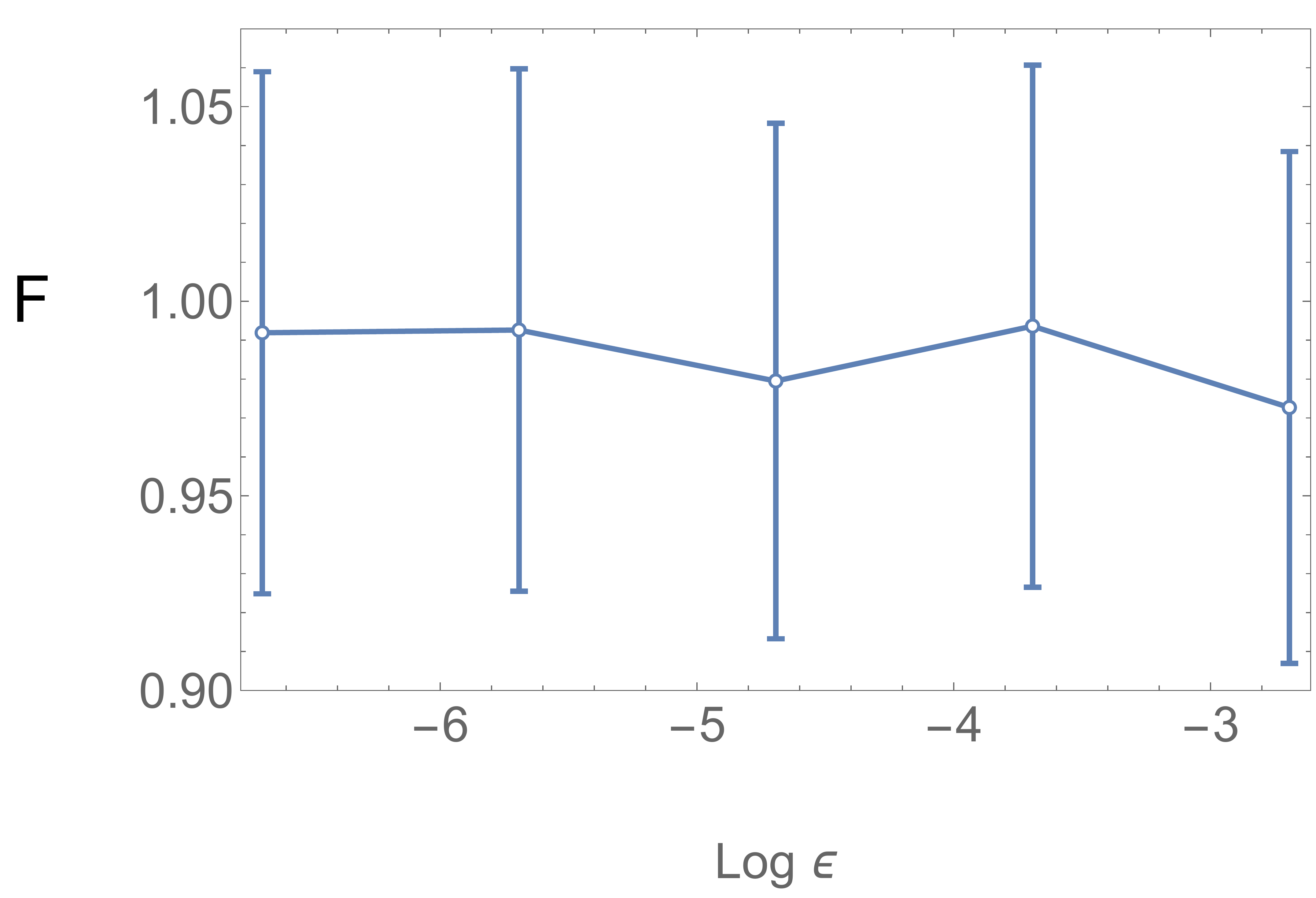} 
 \caption{Fidelities -- approximated by Eq. (\ref{eq:fidapprox})-- vs $\log\varepsilon$ of the target sate with respect to the state in Eq. (\ref{eq:statepert}) for the digital quantum simulation in Auckland with $\varepsilon$ ranging from $0.5\cdot10^{-6}$ to $0.5\cdot10^{-2}$}
    \label{fig:3} 
\end{figure*}

\section*{Conclusions and discussion}

We present results of the digital quantum simulation of a double single-mode squeezing hamiltonian, which is expected to be involved in the generation of quantum entanglement between two quantum harmonic oscillators by a gravitational field. We use a boson-qubit mapping in order to translate bosonic hamiltonians into multiqubit gates. Then we apply gate-decomposition techniques to express them as a sequence of single-qubit and CNOT gates to launch the circuits in IBM quantum devices. We make use of error mitigation strategies -only in the measurement stage- and also post-selection in order to achieve high-fidelity simulations. We choose Auckland as a platform with high QV and the right connectivity. The achieved fidelities are above 90 \% for a large range of parameters, lying in the perturbative regime. Therefore, we generate quantum entanglement by means of an interaction which simulates the action of a quantum gravitational field. While the current experimental proposals develop \cite{Bose:2017nin,Marshman:2019sne,Marletto,interf,newton,versus,constraints,julen,PhysRevD.105.106028}, this quantum simulation platform could inform them by enabling access to parameter regimes that are beyond direct experimental reach. Moreover, the digital approach allows to consider the benefits of quantum error correction techniques \cite{qec}. This could be necessary when considering extensions to higher number of modes or excitations or non-perturbative couplings. With the techniques developed in \cite{sabindqs,paula} it would be straightforward to address the extension to higher number of modes or allowed excitations. For instance, considering that state-of-the art quantum computers contain around 100 superconducting qubits, we could consider up to around 50 maximum excitations per mode, restricting ourselves to two harmonic oscillators. The extension of the operators \cite{sabindqs,paula} would translate into the appearance of interaction terms with more than 4 qubits, which would require an increase in connectivity beyond the current IBM designs  based on heavy-hexagon geometries \cite{heavyhex}. Otherwise, SWAP operations should be introduced to move the forbidden CNOTs to other pair of qubits where the gate is available, therefore increasing significantly the number of gates. We could also add more terms to the perturbative approximation to the state and the fidelity, or even abandon the perturbative approach and go for a full tomography of the state \cite{paula}, which would also increase the number of gates. At this point, as expected for such a large number of qubits, the problem would be presumably hard to simulate with a classical computer \cite{supre,strongsupre}, which illustrates  the necessity of a quantum setup such as the one presented here.  While at the moment we do not have access to a large quantum computer in which all the above could be considered while keeping a good fidelity, the few-qubits high-fidelity circuit reported in this work can be understood as a promising first step along that direction, since the full simulation would consist basically on iterations of the basic building blocks introduced here. Given the rapid development of quantum computers in the last years, we anticipate that in the short term there will be a quantum computer with several hundreds of qubits and improved gate fidelities in which an algorithm consisting in the application of the recipe presented here to more modes and excitations would generate an entangled state with high fidelity. While direct experiments are not available this would be the only way of proving that entanglement can be generated by gravitational means in a quantum setup, since we do not know of any analogue quantum simulation of this system.


\begin{backmatter}

\section*{Declarations}

\section*{Acknowledgements}
I would like to thank Borja Peropadre for invaluable discussions and insightful comments and Paula Cordero Encinar and Andrés Agustí Casado for paving the way of this work.

\section*{Funding}
I acknowledge financial support through the Ramón y Cajal Programme (RYC2019-028014-I).

%
\section*{Availability of data and materials}
Not applicable.
\section*{Ethics approval and consent to participate}
Not applicable.
\section*{Competing interests}
The authors declare that they have no competing interests.

\section*{Consent for publication}
Not applicable

\section*{Authors' contributions}
Not applicable.

\end{backmatter}
\end{document}